\documentclass[12pt]{iopart}

\usepackage{lineno}
\usepackage{graphicx}
\usepackage{subcaption}
\usepackage{amssymb}
\usepackage{xcolor}
\usepackage{color}
\graphicspath{{figures/}}

\begin{document}

\title{A quantum ticking self-oscillator using delayed feedback}

\author{Y. Liu$^{1,3}$, W. J. Munro$^{2}$, and J. Twamley$^{1}$}

\address{$^{1}$Quantum Machines Unit, Okinawa Institute of Science and Technology Graduate University, Onna-son, Okinawa 904-0495, Japan}
\ead{yaananliu@gmail.com; jason.twamley1@oist.jp}
\address{$^2$NTT Basic Research Laboratories \& NTT Research Center for Theoretical Quantum Physics, NTT Corporation, 3-1 Morinosato-Wakamiya, Atsugi-shi, Kanagawa, 243-0198, Japan}
\address{$^3$Centre for Quantum Dynamics, Griffith University, Nathan, QLD 4111, Australia}

\begin{abstract}
Self-sustained oscillators (SSOs) is a commonly used method to generate classical clock signals and SSOs using delayed feedback have been developed commercially which possess ultra-low phase noise and drift. Research into the development of quantum self-oscillation, where one can also have a periodic and regular output {\em tick}, that can be used to control quantum and classical devices has received much interest and quantum SSOs so far studied suffer from phase diffusion which leads to the smearing out of the quantum oscillator over the entire limit cycle in phase space seriously degrading the system's ability to perform as a self-oscillation. In this paper, we explore quantum versions of time-delayed SSOs, which has the potentials to develop a ticking quantum clock. We first design a linear quantum SSO which exhibits perfect oscillation without phase diffusion. We then explore a nonlinear delayed quantum SSO but find it exhibits dephasing similar to previously studied non-delayed systems. 
\end{abstract}
\section{Introduction}
Clocks, one of humanity’s oldest inventions, are arguably one of the most important in today's society having driven its technological revolution. One common method to generate clock signals is using self-sustained oscillators (SSOs), which can be used in many digital circuits and information processing machines, for the purpose of time-keeping and frequency control within different parts \cite{fong2014microwave}. SSOs have been proposed and studied in many different fields, such as mechanical engineering, acoustics, electronics, and even bio-mechanics \cite{Jenkins2013Self-oscillationb,hao2020recent}. Such oscillators with tunable frequencies and very low phase noise are always required in demanding applications \cite{hao2020recent}. 
One example of a tunable high-precision self-sustained oscillator is an optoelectronic oscillator (OEO) which contains a high Q-factor optical resonator and a long optical fiber delay line. This OEO produces oscillation signals with ultra-low phase noise \cite{maleki2011optoelectronic}. Accordingly, building a quantum self-oscillator has also drawn considerable interest in recent decades, since it will play a necessary role in quantum machines. In this paper we will show that a linear quantum self-oscillator without phase diffusion can be achieved by using delayed feedback in optics.

A crucial element of any useful clock is connecting the periodic oscillating system to the outside world so that the {\em tick} of such a clock can be utilised. 
Almost invariably clocks display some limit in phase coherence, either through thermal noise from coupling to a thermal environment, or quantum noise even when coupled to a zero-temperature environment. Most previous studies of self-oscillators considered nonlinear classical or quantum systems whose open system driven dynamics exhibited periodic limit cycles. In this work we consider incorporating time delay, with gain, to produce a self-oscillating system.  We consider cases where one has a linear and then a non-linear open quantum system with time-delay and gain. We can find examples in both the linear and nonlinear systems where the corresponding classical system exhibits closed cycles in phase space.  Similar to previous studies, we find that the  nonlinear quantum dynamics exhibits decaying oscillations of the expectation values which is indicative of phase diffusion over the closed cycles. We also observe that the overall energy in the quantum dynamics is bounded in time. However, the linear quantum dynamics displays periodic oscillations of the expectation values which never decay in time, the state cycles around the closed cycle with no-apparent phase diffusion, but we also observe that the system's energy grows in time.  This growth of the mean energy is  predicted by the thermodynamics of clocks \cite{Milburn2020TheClocks}.

Quantum self-sustained oscillations have primarily focused on the exploration of continuously driven nonlinear quantum systems. These studies include the synchronization of the oscillator with the phase of the drive \cite{Walter2014QuantumOscillator}, as well as the synchronization of two self-oscillators \cite{Lee2013QuantumIons}. These SSOs include nonlinear systems such as the quantum van der Pol oscillator, \cite{Lee2013QuantumIons,Walter2014QuantumOscillators,Lorch2016GenuineSelf-Oscillators,Dutta2019CriticalOscillator}, quantum Rayleigh oscillator \cite{Chia2020RelaxationMechanics}, optomechanical oscillators \cite{Rodrigues2010AmplitudeResonator,Qian2012QuantumInstability}, and atomic oscillators \cite{Xu2014SynchronizationAtoms}. 
Experimental demonstrations of quantum van der Pol oscillators has received little attention with possible experimental implementations using trapped ions \cite{Lee2013QuantumIons}. Quantum optomechanical self-sustained oscillators and atomic oscillators mainly follow the idea from a class of classical electronic SSOs, which is a hybrid of a mechanical(atomic) resonator and a feedback loop with an optical amplifier \cite{fong2014microwave}. 

In all previous studies researchers engineered quantum self-sustained oscillation using a continuous wave driven nonlinear quantum system. In this work we however will examine a very large class of classical SSOs which are formed via a delayed feedback loop and ask if there are quantum versions of such classical delayed SSOs? 
In many works researchers consider the feedback delay to be so small that it can be ignored and the feedback is Markovian. However, substantial feedback delay can be useful, for example, it has been applied in optoelectronic systems to generate high-quality self-oscillators with extremely low phase noise \cite{hao2020recent,Zhang2020NovelRadar,maleki2011optoelectronic}. 
Integrated OEOs have been commercially developed for applications in the fields of communications, radar, signal processing and remote sensing \cite{hao2020recent}. Time-delay in quantum systems have been found to be useful for some quantum information technologies, such as squeezing enhancement and entanglement creation and control \cite{nemet2016enhanced,Hein2015EntanglementFeedback}. Therefore, it will be interesting to consider how to formulate a quantum open system with substantial feedback delay and gain and whether periodic motions can be sustained.


Our paper is structured as follows: we will first consider the design of a delayed linear quantum self-oscillator, which is inspired by a classical linear delayed differential equation (DDE). As the delayed SSOs will require the delayed feedback to be amplified, we reexamine the general quantum cascaded theory for delayed systems and extend this formalism to include delayed amplified feedback. Rather than using matrix-product-states to tackle the growing Hilbert space dimension, we develop mean-field approximations to enable quantum simulations for longer times.  
We first study the open dynamics of a linear quantum delayed oscillator with gain and we show that by correctly matching the natural period of the oscillator with no feedback, with the period of the self-sustained delay/gain oscillations, one can obtain a large family of classical closed cycles in phase-space. We are able to use our mean-field method to show that the full quantum dynamics follows this classical cycle for several delay periods with no evidence of decay.  
We then explore a class of nonlinear delayed quantum self-oscillators, where we introduce two-photon absorption in the optical cavity where the dynamics are saturated naturally by the nonlinearities. The inclusion of a non-linearity is far more challenging to model even using the mean-field methods as the number and complexity of the differential equations grows dramatically. We perform the same analysis as in the linear case and in all situations studied the oscillations decay. We expect therefore, that we have strong evidence that in our linear delayed-SSO the quantum dynamics follows a closed cycle with no phase diffusion but the energy grows with time, while in the nonlinear delayed-SSO the energy is bounded and the system appears to exhibit phase diffusion.  Phase diffusion in the quantum dynamics could be more explicitly studied by using the Wigner function but the numerical cost to simulate quantum dynamics whose extent is large enough to observe a clearly identified closed cycle in the Wigner function is prohibitive.  

\section{Linear quantum self-oscillators driven by time-delay feedback}
\label{sec:linear}
\begin{figure}
    \centering
    \includegraphics[width=\textwidth]{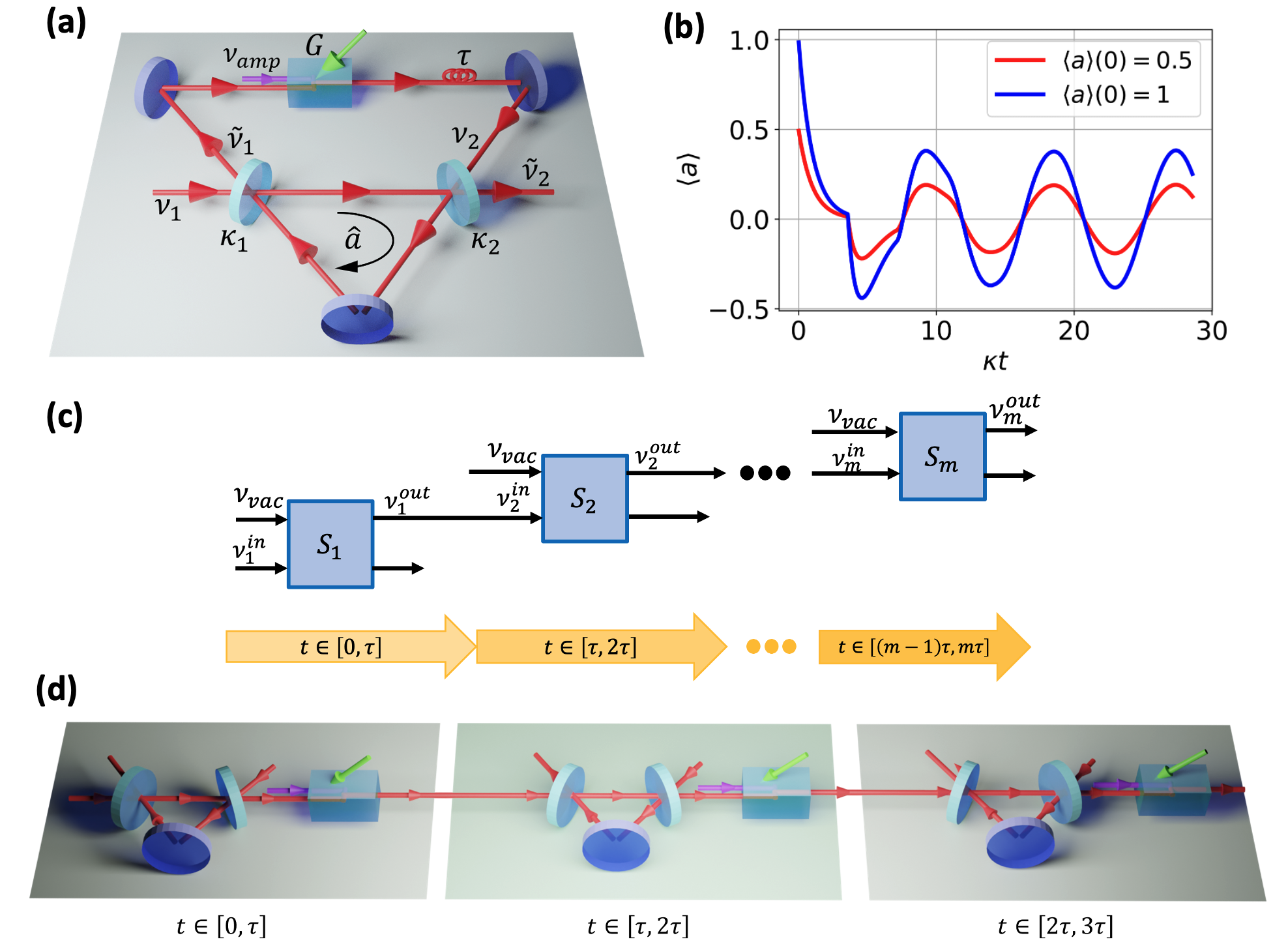}
\caption{(a) The ring cavity is composed by two partially reflective mirrors (in light blue) and one fully reflective mirror (in light purple) confines a standing mode $\hat{a}$. One output field $\tilde{\nu}_1$ is fed back to the cavity after amplification $G$ and a delay time $\tau$; the green arrow towards to the amplifier represents a pumping signal; (b) The self-oscillation of $\langle \hat{a} \rangle$ for two different initial values in linear quantum optical system, solved by using mean field approximation. $\langle \hat{a} \rangle$ shows perfect self oscillation; (c) Pictorial representation of the cascaded theory of a quantum system with time-delay feedback; each system $S_j$ in the chain represents evolution in the time interval $\left[(j-1)\tau,\tau\right]$, the current subsystem is driven by output from previous subsystem in time $t-\tau$; (d) The cascaded chain of the optical quantum system with time-delay shown over three time intervals; 
}
\label{fig:1}
\end{figure}

In this section we first consider a linear quantum system with feedback and gain, whose Langevin equations will be described by DDEs. The system we consider is shown schematically in Fig. \ref{fig:1}-(a). Note that we use this figure to better visualise the space-time conncectivities and it is not a proposal for a real experiment. We consider a ring cavity formed by two partially reflective mirrors and one fully reflective mirror which confines a circulating mode of light $\hat{a}$. We arrange so that a portion of the light from this ring cavity will exit the cavity, pass through an amplifier with gain $G$, and re-enter the cavity after a delay time $\tau$.  

In the following we first present the Langevin equations for our system, incorporating time delay and gain, and find that the form of these equations are identical to classical DDEs \cite{Smith2010AnSciences}. Subsequently, we investigate the properties of such classical DDEs to identify an infinite set of parameters that yield periodic oscillations in their solutions. Next, we derive the master equation for these time-delayed quantum systems using cascade theory \cite{Gardiner2004QuantumOptics}, and present simulation results.  We observe that the dynamics of $\langle \hat{a}\rangle$ inside the cavity exhibit oscillation without decay - i.e. forever.  We then consider the system (without feedback), to be a harmonic oscillator. When we implement delayed feedback on this oscillator we can find values for the frequency of this oscillator that result in closed cycles in phase space.  In these cases the mean quantum dynamics $(\langle x(t)\rangle,\langle p(t)\rangle)$, follows this closed cycle and we show this numerically for the full quantum dynamics over the first few delay periods.

\subsection{Langevin equation of the cavity with time-delay feedback and gain}
First studied in 1994 \cite{Wiseman1994QuantumFeedback}, time-delay in quantum feedback has not enjoyed much progress primarily due to the significant challenge faced to obtain a valid master equation. When the delay time is not infinitesimal, the quantum dynamics becomes non-Markovian \cite{Grimsmo2015Time-DelayedControl}, which can be quite problematic. There are, to date only a few methods discovered to describe a quantum master equation which incorporates time delayed feedback. These methods involve the use of cascaded quantum channels and can use matrix product states (MPS) to help cope with the very large Hilbert spaces that appear \cite{Grimsmo2015Time-DelayedControl,Pichler2016PhotonicFeedback}. By considering the time-delayed quantum system in a cascaded way, where the current system is driven (affected) in a temporally directed fashion by its past dynamics, a master equation in piece-wise form can be derived \cite{Whalen2017OpenFeedback}. In this section, we utilise and generalise the cascaded theory to get the master equation for quantum self-oscillator with time delay.

We can first write the Langevin equation of the cavity mode $\hat{a}$ as:
\begin{equation}
\label{eqn:Langevin_a}
\eqalign{
        \dot{\hat{a}}(t)&=-\kappa \hat{a}(t)-\sqrt{\kappa_1}\nu_1(t)-\sqrt{\kappa_2}\nu_2(t),\cr
        \tilde{\nu}_1(t)&=\sqrt{\kappa_1}\hat{a}(t)+\nu_1(t),\\
        \tilde{\nu}_2(t)&=\sqrt{\kappa_2}\hat{a}(t)+\nu_2(t),}
\end{equation}
where $\nu_1(t)$ ($\nu_2(t)$) are the input fields of the left (right) partial mirrors with $\tilde{\nu}_1(t)$ ($\tilde{\nu}_2(t)$) being the output fields of left (right) partial mirrors. The decay rates of these mirrors are $\kappa_1$ ($\kappa_2$) respectively, with $\kappa=\frac{ \kappa_1+\kappa_2 }{2}$ being the average decay of the cavity. 

We insert a quantum amplifier into the feedback loop, so we get $G>1$. It is well known that such amplification cannot be achieved without incorporating noise and continuous-wave driving \cite{josse2006universal}. We can express the returning feedback signal $\nu_2(t)$ following a delay, and amplification on the signal $\tilde{\nu}_1(t-\tau)$, as:
\begin{equation}
    \label{eqn:feedbackloop}
    \nu_2(t)=e^{i\phi}\left( \sqrt{G}\,\tilde{\nu}_1(t-\tau)+\sqrt{G-1}\,\nu_{amp}^\dagger(t-\tau) \right),    
\end{equation}
where $\phi$ is the phase change in the feedback loop while $\nu_{amp}$ is the amplifier noise field \cite{josse2006universal}. Substituting \Eref{eqn:feedbackloop} into \eref{eqn:Langevin_a}, we obtain
\begin{equation}
\label{eqn:QuantumDDEsto} 
\eqalign{
   \dot{\hat{a}}(t)&=-\kappa \hat{a}(t)-e^{i\phi} \sqrt{G\kappa_1\kappa_2}\,\hat{a}(t-\tau)-\sqrt{\kappa_1}\,\nu_1(t)-\sqrt{G\kappa_2}\,\nu_1(t-\tau)\\
   &-e^{i\phi}\sqrt{(G-1)\kappa_2}\,\nu_{amp}^\dagger(t-\tau).
}
\end{equation}
The noise field of the amplifier appears in the form of $\nu_{amp}^\dagger$ in \eref{eqn:QuantumDDEsto}, which will inject  thermal noise into the system due to the relations  $d\nu_{amp}^\dagger(t) d\nu_{amp}(t)=\bar{N}_{amp}dt$ and $d\nu_{amp}(t)d\nu_{amp}^\dagger(t) =(\bar{N}_{amp}+1)dt$ (see \ref{apx:Qamplification} for details).
If we assume the input field of the left mirror $\nu_1(t)$ and the quantum amplifier noise field $\nu_{amp}(t)$ are all vacuum fields, taking the expectation values of the above quantum DDE \eref{eqn:QuantumDDEsto} leads to:
\begin{equation}
\label{eqn:QuantumDDE}
    \langle \dot{\hat{a}}(t) \rangle=-\kappa \langle \hat{a}(t) \rangle-e^{i\phi}\sqrt{G\kappa_1\kappa_2}\,\langle  \hat{a}(t-\tau)\rangle.
\end{equation}
This quantum DDE has exactly the same form to the following general classical DDE \cite{Smith2010AnSciences}
\begin{equation}
\label{eqn:linear_dde}
    \dot{x}(t)=\alpha x(t)+\beta x(t-\tau),
\end{equation}
where $\alpha\in \mathbb{R}, \beta\in \mathbb{R}$ are the usual system parameters.

This classical DDE has been extensively explored \cite{Brunovsky2004OnSolutions} and we will investigate its properties below. In particular, we will obtain conditions on the system parameters $\alpha,\beta$, and $\tau$ to obtain a self-sustained oscillation of $x(t)$, and hence $\langle \hat{a}(t) \rangle$ in \Eref{eqn:QuantumDDE}.

\subsection{Classical time-delay differential equations}
Systems with time-delay feedback are usually described using DDEs in the form of \Eref{eqn:linear_dde}. Such DDEs have been widely studied in various fields such as population biology \cite{May2001StabilityEcosystems}, physiology \cite{Glass1988FromChaos}, epidemiology \cite{Busenbergf1978PeriodicEquation}, economics \cite{Brunovsky2004OnSolutions}, neural networks \cite{VanDenDriessche2001StabilizationNetwork}, and electronic circuits \cite{Bellen1999MethodsType} and thus provide an ideal basis to explore SSOs.

\begin{figure}
    \centering
    \includegraphics[width=0.95\textwidth]{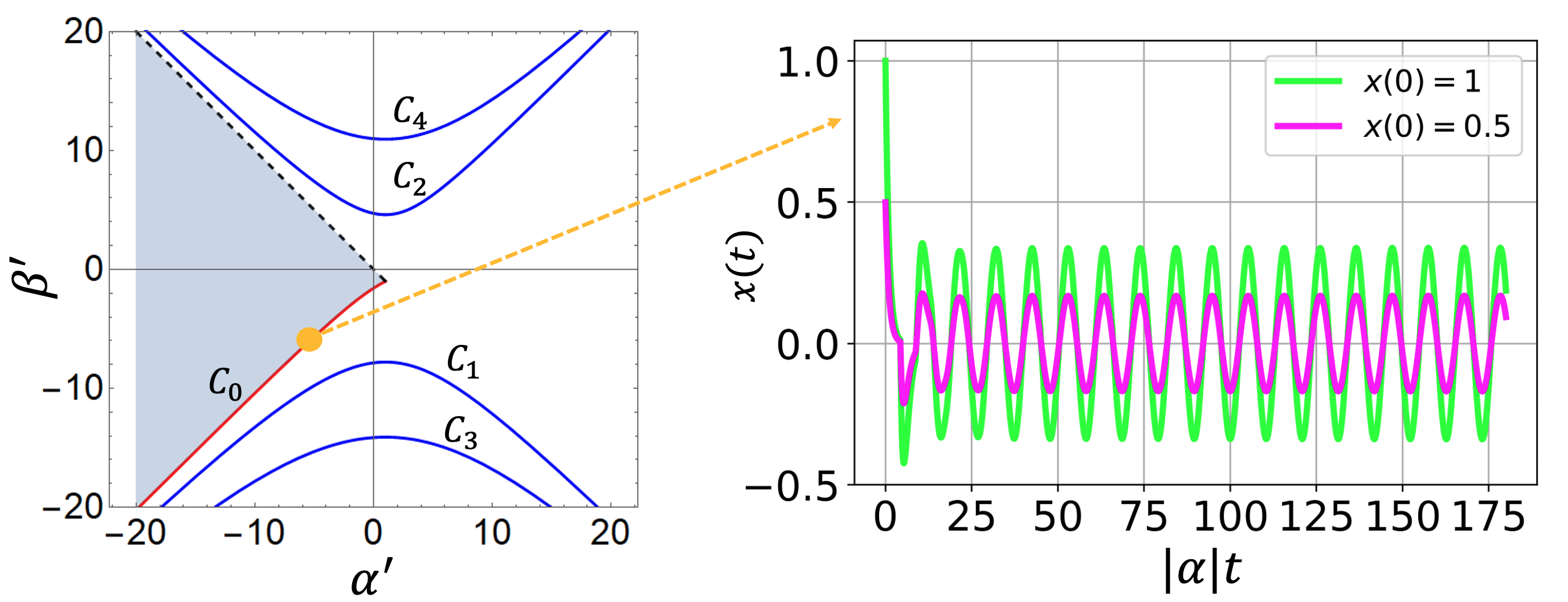}
    \caption{Illustration showing the self-oscillation for the linear DDEs; the left figure is the parameters space $(\alpha^\prime,\beta^\prime)$ showing the (shadow) stable region $\Re{z}<0$, unstable (white), and red and blue curves $C_0, C_1, C_2, \cdots$. We consider the Red curve $C_0$ and $(\alpha^\prime,\beta^\prime)\in C_0$ show perfect SSO; the right figure shows two example solutions $x(t)$ with $(\alpha^\prime,\beta^\prime)\in C_0$ indicating self-oscillation. The amplitude of oscillation depends on the initial value but the oscillations have identical periods and exhibit no decay or phase noise/diffusion.}
    \label{fig:2}
\end{figure}

If we assume the solution of the \Eref{eqn:linear_dde} has the exponential form $x(t)=e^{\lambda t}$, then substituting this solution into \Eref{eqn:linear_dde} gives us the characteristic equation (CE), $\lambda-\alpha -\beta e^{-\lambda \tau}=0$ which we can rewrite as
\begin{equation}
\label{eqn:chareqn}
    z-\alpha^\prime-\beta^\prime e^{-z}=0,
\end{equation}
where we have multiplied the CE by $\tau$, and denote $z=\lambda \tau, \alpha^{\prime} =\alpha\tau, \beta^{\prime} =\beta\tau$. The solutions of \Eref{eqn:chareqn} for the characteristic values of $z$ can be written as $z=\Re{z}+i\Im{z}$, which means that the solution to \Eref{eqn:linear_dde} is $x(t)=\textrm{Exp}\left[\frac{\Re{z}}{\tau}t\right] \textrm{Exp}\left[i\frac{\Im{z}}{\tau}t\right]$. The behavior $\textrm{Exp}\left[\frac{\Re{z}}{\tau}t\right]$ describes whether $x(t)$ is increasing or decaying based on the sign of $\Re{z}$, while latter $\textrm{Exp}\left[i\frac{\Im{z}}{\tau}t\right]$, describes oscillations of $x(t)$.
With this notation, \Eref{eqn:chareqn} becomes:
\begin{equation}
\label{eqn:chareqn_div}
\eqalign{
\Re{z}-\alpha^{\prime}-\beta^{\prime} e^{-\Re{z}}\cos{\Im{z}}&=0,\\
\Im{z}+\beta^{\prime} e^{-\Re{z}}\sin{\Im{z}}&=0.}
\end{equation}
To ensure $x(t)$ oscillates, we need to guarantee that \Eref{eqn:chareqn} has purely imaginary solutions, that is, $\Re{z}=0$. The values of $(\alpha^\prime,\beta^\prime)$ satisfying this fall into separate classes depending on the argument of $z$, with the parameters $(\alpha^\prime,\beta^\prime)$ in sets of one-dimensional curves, which we denote as $C_j$ with 
\begin{equation*}
\left\{
\eqalign{
    (\alpha^\prime,\beta^\prime)\in C_0&, \Im{z}\in (0,\pi) \\
    (\alpha^\prime,\beta^\prime)\in C_1&, \Im{z}\in (\pi,2\pi)\\
    &\vdots \\
    (\alpha^\prime,\beta^\prime)\in C_j&, \Im{z}\in (j\pi,(j+1)\pi)}\right.
\end{equation*}
as shown in \Fref{fig:2}. We consider $(\alpha^\prime,\beta^\prime)\in C_0$ from here on and these solutions display the greatest level of robustness. In the right of \Fref{fig:2} we plot the SSO of \Eref{eqn:linear_dde} for $(\alpha^\prime,\beta^\prime)=(-3.573,-4.375)$,
where we consider two initial states, $x(t=0)=1$ and $x(t=0)=0.5$ with $x(t<0)=0$. We observe that by choosing parameters on the line $C_0$ we can obtain perfect self-oscillation for all time where the oscillation amplitude depends on the initial value.

\subsection{Master equation of quantum self-oscillators}
From studying the classical DDEs \eref{eqn:linear_dde}, we find a one-dimensional set of parameters $(\alpha^\prime,\beta^\prime)$ that displays self-sustained oscillation in $x(t)$. When $(\alpha^\prime, \beta^\prime)\in C_0$, we can always find a critical delay time 
$\tau_{cr}=\arccos(-\frac{\alpha}{\beta})/\sqrt{\beta^2-\alpha^2}$
to achieve self-oscillation if $|\beta|>|\alpha|$. Returning now to the quantum expectation in \Eref{eqn:QuantumDDE} for $\langle \hat{a}(t) \rangle$, we observe that in the case $\phi=0$, for any set of quantum parameters $(\kappa_1, \kappa_2, G)$, we can find a critical delay $\tau_{cr}$ such that the quantum system has pure self-oscillation in $\langle \hat{a}(t)\rangle$. The field will oscillate forever, with the only the CW drive required by the quantum amplifier $G$, as long as we can guarantee the delay time is set as $\tau_{cr}$.

Although the time-delayed quantum Langevin equations give an elegant description of the system, it is unclear how to solve them numerically. An alternative method is to derive a time-delayed master equation which can be solved numerically. One approach to derive such a master equation with time-delayed feedback is based on cascaded channel theory \cite{Gardiner1993DrivingSystem,Carmichael1993QuantumSystems}. This has been previously been applied to model quantum systems with time-delayed feedback when amplification is absent \cite{Whalen2017OpenFeedback}. The basic idea is to copy the quantum system into multiple identical new systems, each of them representing the evolution of the original system only within a time interval $\tau$ - the delay time \cite{Grimsmo2015Time-DelayedControl}. This we schematically depict in \Fref{fig:1}-(b), where $S_1, S_2, \cdots, S_m$ are $m$ copies of the system. For $t\in [0,\tau]$, only the first copy $S_1$ evolves, which represents the evolution of the original system before the delayed feedback signal appears. When $t\in [\tau,2\tau]$, the second copy $S_2$ is involved to represent the evolution of current system, and the delayed feedback signal will be represented as the output of the previous copy $S_1$. Similarly, we can use a new copy $S_i$ to represent the most current system, which will be driven by its previous copy. By writing the master equation of this cascaded chain, we can derive the master equation of the time-delayed quantum system. However, the cascaded theory in \cite{Gardiner2004QuantumOptics} needs to be extended before we can apply it to our case. The reason is that we consider an amplifier in the feedback loop and this quantum amplifier noise that effects subsequent temporal dynamics will be amplified step by step.

To obtain such a master equation, we first start from the coupling Langevin equation. Then by defining all the input fields for each system in the cascaded chain, we can derive the Ito white noise quantum stochastic differential equation as:
\begin{equation}
\label{eqn:ito_equation}
\eqalign{
        d\hat{o}_m=-\frac{i}{\hbar}[\hat{o}_m,H_{\textrm{tot}}]dt-\sum_{j=1}^m  \frac{\gamma_j}{2}(\bar{N}_j+1)\left( 2\hat{a}_j^\dagger \hat{o}_m\hat{a}_j-\hat{o}_m\hat{a}_j^\dagger \hat{a}_j-\hat{a}_j^\dagger \hat{a}_j \hat{o}_m \right)dt\\
        -\sum_{j=1}^m\frac{\gamma_j}{2}\bar{N}_j\left( 2\hat{a}_j\hat{o}_m\hat{a}_j^\dagger-\hat{o}_m\hat{a}_j\hat{c}_j^\dagger-\hat{a}_j\hat{a}_j^\dagger\hat{o}_m \right)dt \\
        +\sum_{j=1}^m\left\{ -[\hat{o}_m,\hat{a}_j^\dagger]\sqrt{G\gamma_{j-1}\gamma_n}\hat{a}_{j-1}+\sqrt{G\gamma_{j-1}\gamma_j}\hat{a}_{j-1}^\dagger [\hat{o}_m,\hat{a}_j] \right\}dt \\
       +\sum_{j=1}^m\{ -[\hat{o}_m,\hat{a}_j^\dagger]\sqrt{\gamma_j}\left(\epsilon_{in}(j,t)dt+dB(j,t)\right)\cr
       +\sqrt{\gamma_j}\left(\epsilon^*_{in}(j,t)dt+dB^\dagger(j,t)\right)[\hat{o}_m,\hat{a}_j] \}.}
\end{equation}
Here we assume there are $m$ systems in the cascaded chain, and $H_{\textrm{tot}}$ is the sum over all system Hamiltonians. The input fields into the $j$-th system in the chain is defined as $db_{j}^{in}(t)=\epsilon_{in}(j,t)dt+dB(j,t)$, where $\epsilon_{in}(j,t)$ is the coherent part and $dB(j,t)$ is the white noise part. We use $\gamma_j$ as the decay rate of the $j$-th system, and all the operators have the form of:
\begin{equation}
    \mathcal{O}_j=\underbrace{I\otimes\cdots \otimes I}_{j\textit{ times}} \otimes \mathcal{O}\otimes \underbrace{I \otimes \cdots \otimes I}_{(m-j-1) \textit{ times}}.
\end{equation}
The operator $\mathcal{O}$ is in the Hilbert space of each individual system, and the operator $\mathcal{O}_j$ resides in the whole Hilbert space which is a tensor product of all the temporally separate systems. The details to obtain \Eref{eqn:ito_equation} can be found in the \ref{apx:mastercalculation}. Using this Ito equation, a master equation can be derived for the density operator $\rho_m(t)$ for the chained system, that is, $\rho_m(t)\in \otimes_{j=1}^{m}\mathcal{H}_S(j)$, where $\mathcal{H}_S(j)$ is the Hilbert space of individual systems in the chain.  
The master equation can be obtained by setting $\langle d\hat{o}(t)\rho_m \rangle =\langle d\hat{o}\rho_m(t) \rangle$, to obtain the evolution of the entire historical chain up to the $m$-th time interval $t\in[(m-1)\tau,m\tau]$:
\begin{equation}
\label{eqn:cascaded_master}
\eqalign{
\frac{d\rho_m}{dt}&=\frac{i}{\hbar}[\rho,H_{\textrm{tot}}]+\sum_{j=1}^m\left\{\gamma_j (\bar{N}_j+1)\mathcal{D}[\hat{a}_j]\rho+\gamma_j \bar{N}_j\mathcal{D}[\hat{a}_j^\dagger]\rho\right\}\\
&+\sum_{j=2}^m\sqrt{G\gamma_{j-1} \gamma_j}\left\{ [\hat{a}_j,\hat{a}_{j-1}\rho]+[\rho \hat{a}_{j-1}^\dagger,\hat{a}_j] \right\}\\
&-\sum_{j=1}^m\left\{\sqrt{\gamma_j}\left[\hat{a}_j^\dagger,\rho  \right]\epsilon_{in}(j,t)-\sqrt{\gamma_j}\epsilon^*_{in}(j,t)\left[\hat{a}_j,\rho\right]\right\}\\}
\end{equation}
with the initial state at $t=0$ being:
\begin{equation}
    \label{eqn:initial_state}
    \rho_m(0)=\underbrace{\rho_S(0)\otimes \cdots \otimes \rho_S(0)}_{m \textit{ times}}.
\end{equation}
This master equation has exactly the same form to the general cascaded theory in \cite{Gardiner2004QuantumOptics}, but with the modified mean ancestor bath occupation of the fields $\bar{N}_j$, the detail of which can be found in \ref{apx:mastercalculation}. If we remove the amplifier, by setting $G=1$, and let the input and amplifier noise fields be vacuum fields, \Eref{eqn:cascaded_master} reduces exactly to that derived in \cite{Whalen2015OpenInteractions}.

Armed with the general cascaded description incorporating amplification, we can now return to our original time-delayed system shown in Fig \ref{fig:1}-(a). Similar to the idea in \Fref{fig:1}-(b), we consider the time-delayed optical system in a cascaded way. This can be shown in \Fref{fig:1}-(c), where we use one copy of the ancestor cavity to represent its evolution in different time intervals of length $\tau$. The output of one copy will be amplified and used to drive the next copy in the chain. 
To numerically simulate this we will use $\hat{a}_0$ to represent the  current system, and $\hat{a}_{m}$ to represent older systems, where larger $m$ means older iterates. The master equation of the time-delayed optical system shown in \Fref{fig:1}-(a) when $t\in [m\tau,(m+1)\tau]$ is thus:
\begin{equation}
\label{eqn:master_equation}
\eqalign{
\frac{d\rho}{dt}&=\mathcal{L}_m\rho, t\in [m\tau, (m+1)\tau], \textrm{with}\\
\mathcal{L}_m\rho&=-\frac{i}{\hbar}\sum_{j=0}^m [H_j,\rho]+\sum_{j=0}^{m}  \bar{N}_{m-j}\kappa_1\left(\mathcal{D}[\hat{a}_j]\rho+\mathcal{D}[\hat{a}_j^\dagger]\rho\right)\\
&+(\kappa_1+\kappa_2)\sum_{j=0}^m \mathcal{D}[\hat{a}_j]\rho -\sqrt{G\kappa_1 \kappa_2}\sum_{j=1}^{m}\left\{ [\hat{a}_{j-1}^\dagger,\hat{a}_j\rho]+[\rho \hat{a}_j^\dagger, \hat{a}_{j-1}] \right\},
}
\end{equation}
where 
\begin{equation}
    \label{eq:meanoffield_our_setup}
    \bar{N}_{m-j}=G\bar{N}+(G-1)(\bar{N}_{amp}+1), \textrm{for } j=0,1,\cdots, m,
\end{equation}
where $\bar{N}$ is the bath occupation of the input field to the initial system, and $\bar{N}_{amp}$ is the bath occupation of the noise input to the amplifier.
If we assume the input field at the initial time interval and the noise input to the amplifier are both vacuum field, this will simplify $\bar{N}_{m-j}$ as:
\begin{equation}
    \label{eqn:mean_field_vacuum}
    \bar{N}_{m-j}=G-1, \textrm{for } j=0,1,\cdots, m.
\end{equation}
Therefore, we can identify $\hat{a}_j$ in \Eref{eqn:master_equation} with the operator $\hat{a}(t-j\tau)$ in the time-delay Langevin equation \eref{eqn:QuantumDDE}, and the expectation value of the most current time step we aim to calculate will correspond to the zeroth system $\langle \hat{a}_0 \rangle $.

\subsection{Behavior of $\langle\hat{a} \rangle$ and $\langle \hat{a}^\dagger\hat{a}\rangle$ in linear quantum self oscillators}
In this section, we will employ two approaches to simulate the evolution of the expectation values for the {\em current system annihilation operator},  $\langle \hat{a}_0\rangle$ and occupation $\langle\hat{a}_0^\dagger \hat{a}_0\rangle$. We observe that the $\langle \hat{a}_0\rangle$ exhibits indefinite oscillation without decay, albeit at the expense of an increasing energy. 

The first approach uses the piece-wise master equation \eref{eqn:master_equation} to obtain $\langle \hat{a}_0 \rangle={\rm Tr}(\hat{a}_0\rho)$. If we truncate the Hilbert space for each system in the chain to be $N_{trunc}$, the total dimension of the Hilbert space of the complete cascaded chain, when one evolves over $m$ delay intervals, grows as $N_{trunc}^m$. The computational cost to follow this grows to be prohibitive even when $m\sim 5-10$. This is the principle computational bottleneck for this approach. Such large Hilbert spaces motivates some researchers to explore using MPS to address this \cite{Pichler2016PhotonicFeedback}. The increasing Hilbert space dimension required becomes quickly very expensive as regards computer storage, and typically we can only simulate for a few time steps. Furthermore, increasing the truncation $N_{trunc}$ will further increase the storage resources. Therefore, we propose a second approach to simulate $\langle \hat{a}_0 \rangle$ for extended durations, which we call the mean-field approach. For a given time interval $t\in [m\tau, (m+1)\tau]$, we first derive a set of coupled ordinary differential equations (ODEs) in terms of $d\langle \hat{a}_j \rangle/dt, j=0,\cdots m$ in each time step from $[0,\tau]$ to $[m\tau,(m+1)\tau)]$. By solving this set of ODEs, we obtain the evolution of the most current system annihilation operator $\langle \hat{a}_0(t)\rangle$. We also wish to observe the evolution of photon number $\langle \hat{a}_0^\dagger \hat{a}_0 \rangle$, with time. For this we will need to obtain sets of ODEs involving product moments between two systems. { Although the number of required ODEs, $N_{Eqns}$,  grows in time like $\sim O( m^2)$,} when integrating over $m$-delay periods, the required storage  is still much smaller than that needed to solve the full quantum master equation. We perform several simulations using both the full quantum master equation and the above mentioned mean-field approach.

\begin{figure*}[!ht]
\centering
\includegraphics[width=\textwidth]{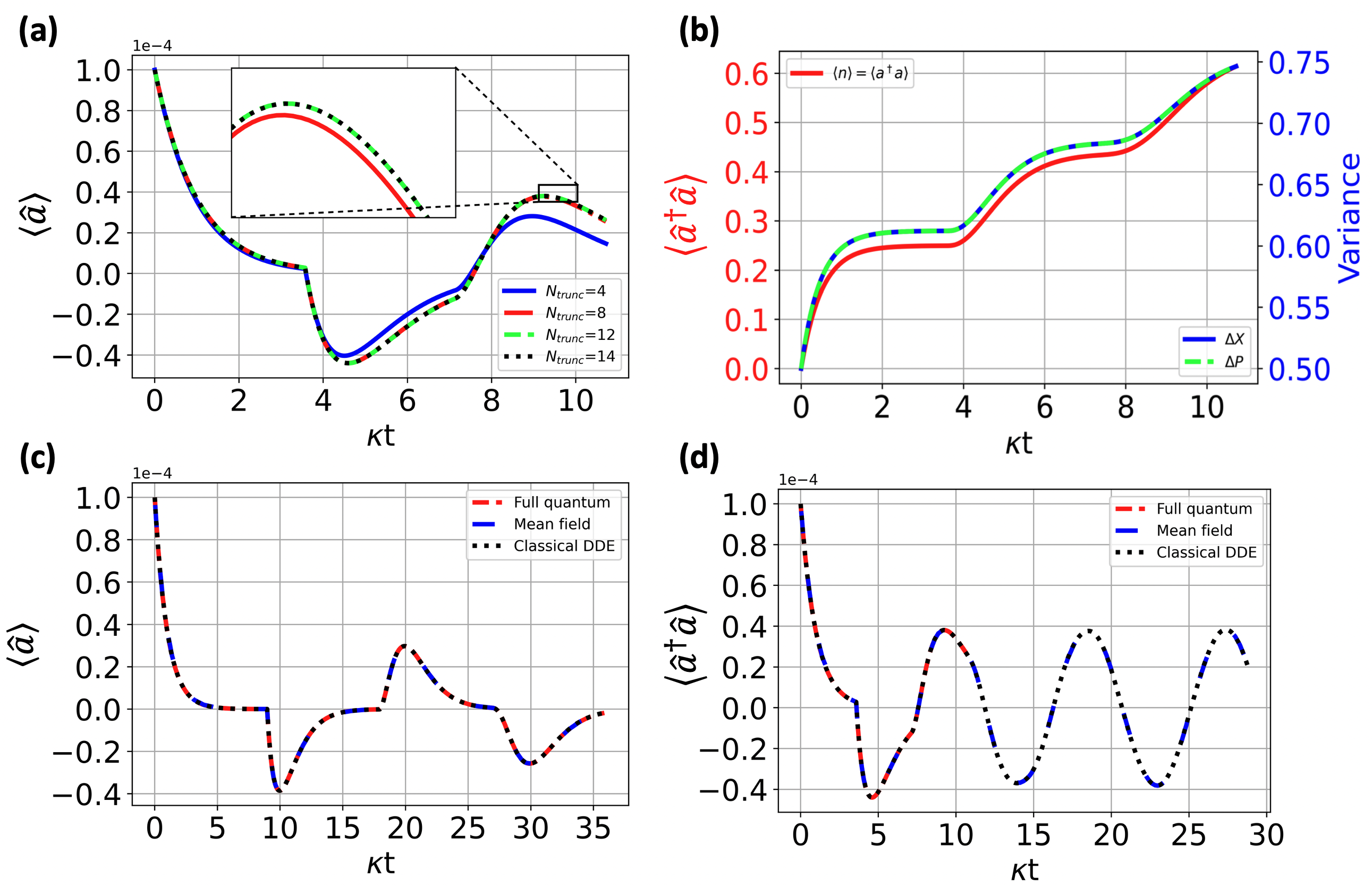}
\caption{Linear self-sustained quantum oscillators show perfect oscillation of $\langle\hat{a}\rangle$ with the cost of unbounded energy increasing. (a) Plots the convergence of $\langle \hat{a} \rangle$ with the Fock space truncation $N_{trunc}=(4,8,12,14)$; the inset shows the zoomed-in detail for $\kappa t\in [9,10]$; (b) Time evolution of $\langle\hat{a}^\dagger \hat{a} \rangle$, $\Delta X(t)$, and $\Delta P(t)$ solved using the full quantum master equation; (c) Oscillating dynamics of $\langle \hat{a}\rangle$ setting the feedback gain $G=1.1$, solved using the full quantum master equation, the mean-field approach, and the classical DDE respectively. Each solution method displays excellent agreement with each other; (d) Oscillating dynamics of $\langle \hat{a}\rangle$ setting the feedback gain $G=1.5$, evolved over 8 time steps ($8\tau$), using the full quantum master equation, the mean-field approach and the classical DDE, displaying excellent agreement.}
\label{fig:4}
\end{figure*}

For simplicity, we can choose the decay rates $\kappa_1=\kappa_2=\kappa$, the feedback gain as $G=1.5$, which gives us $\alpha/\kappa=-1, \beta/\kappa=-1.225$, and $\kappa\tau\approx 3.592$. 
This gives the self-oscillation frequency of $\langle \hat{a}_0 \rangle$ as 
$\omega_\tau/\kappa\approx 0.707$. 
Due to the gain $G>1$ in the feedback loop, we need to choose a suitable truncation of the Fock space and we explore the convergence of our simulations as a function of this truncation in \Fref{fig:4}-(a). 

From \Fref{fig:4}-(a) we observe that $\langle \hat{a}_0\rangle $ shows oscillation in time due to the feedback gain starting from the second time interval. We also observe good convergence if we choose a Hilbert space truncation $N_{trunc}\geq12$. Choose $N_{trunc}=12$ we evolve the photon number $\langle n \rangle=\langle \hat{a}_0^\dagger \hat{a}_0 \rangle$, as well as the variance of the amplitude quadrature $X$ and phase quadrature $P$, calculated as $\Delta X=\sqrt{\langle X^2\rangle-\langle X \rangle^2}$ and $\Delta P=\sqrt{\langle P^2 \rangle-\langle P\rangle^2}$. The results are shown in \Fref{fig:4}-(b), which indicates that the photon number and the quadrature variance increase with time. We have no reason to expect that this behaviour will saturate in time, i.e. we expect the energy of the system to grow indefinitely.

{\subsection{Linear dynamics exhibiting closed cycles in phase space}
In \Fref{fig:4}, $\langle a\rangle$ exhibits sustained oscillation without any decay, which strongly suggests the absence of phase diffusion in  phase space. For the delayed oscillator described by the linear DDE  in \Eref{eqn:linear_dde}, the trajectories in phase space $(\langle x(t)\rangle, \langle p(t)\rangle)$, actually describe straight lines which is not typically what one considers to be a closed cycle. By modifying the system slightly we can devise periodic dynamics which do exhibit closed curves in phase space more clearly.  To do this we include a non-zero frequency term in the Hamiltonian of the non-delayed system, which results in the following modified DDE:
\begin{equation}
\label{eqn:linear_dde_omega}
    \dot{x}(t)=i\omega x(t)+\alpha x(t)+\beta x(t-\tau)\;\;.
\end{equation}
We can actually find new coordinates to relate (\ref{eqn:linear_dde_omega}), to the original DDE (\ref{eqn:linear_dde}). To do this we multiply both sides of (\ref{eqn:linear_dde_omega}), by $e^{-i\omega t}$, and let $\tilde{x}(t)=e^{-i\omega t}x(t)$, and then we can rewrite the DDE for $\tilde{x}$ as:
\begin{equation}
\label{eqn:linear_dde_omega_tilde}
    \dot{\tilde{x}}(t)=\alpha \tilde{x}(t)+\beta e^{-i\omega \tau} \tilde{x}(t-\tau),
\end{equation}
with complex coefficient $\beta^\prime=\beta e^{-i\omega \tau}$. By choosing $\omega\tau=2n\pi$ for an integer $n$, we can have $\beta^\prime=\beta$. In this scenario a closed curve cycle, (rather than a straight line), can be observed between the quadratures $x(t)$ and $p(t)$ when starting from a complex initial state. 

We present simulation results for these closed curved  cycles in \Fref{fig:closedcycles}, for the mean quadratures either by solving their corresponding DDEs \Eref{eqn:linear_dde_omega}, or by a full quantum simulation of the delayed dynamics. 

In \Fref{fig:closedcycles}-(a), we solve \Eref{eqn:linear_dde_omega} for two different different closed cycles corresponding to two values of $\beta$,for the same initial point. While in  \Fref{fig:closedcycles}-(b) we plot the solution of the DDE and the mean values of the quadratures $(\langle x(t)\rangle, \langle p\rangle)$, for the full delayed master equation. For the DDE we evolve forward in time from $t=0$, but we also have to specify the values of the dependent variables for $t<0$, and by exponentially damping them to zero $x,p\sim \exp(\gamma t),\;\;t<0$, we find a very good fit for the full quantum evolution over the first two delay time steps.
The trajectories in phase space converge to their corresponding closed cycles. Notably, for the linear DDE, the closed cycles are influenced by decay rates, feedback gain, and initial states.

\begin{figure}[!ht]
\centering
\includegraphics[width=\textwidth]{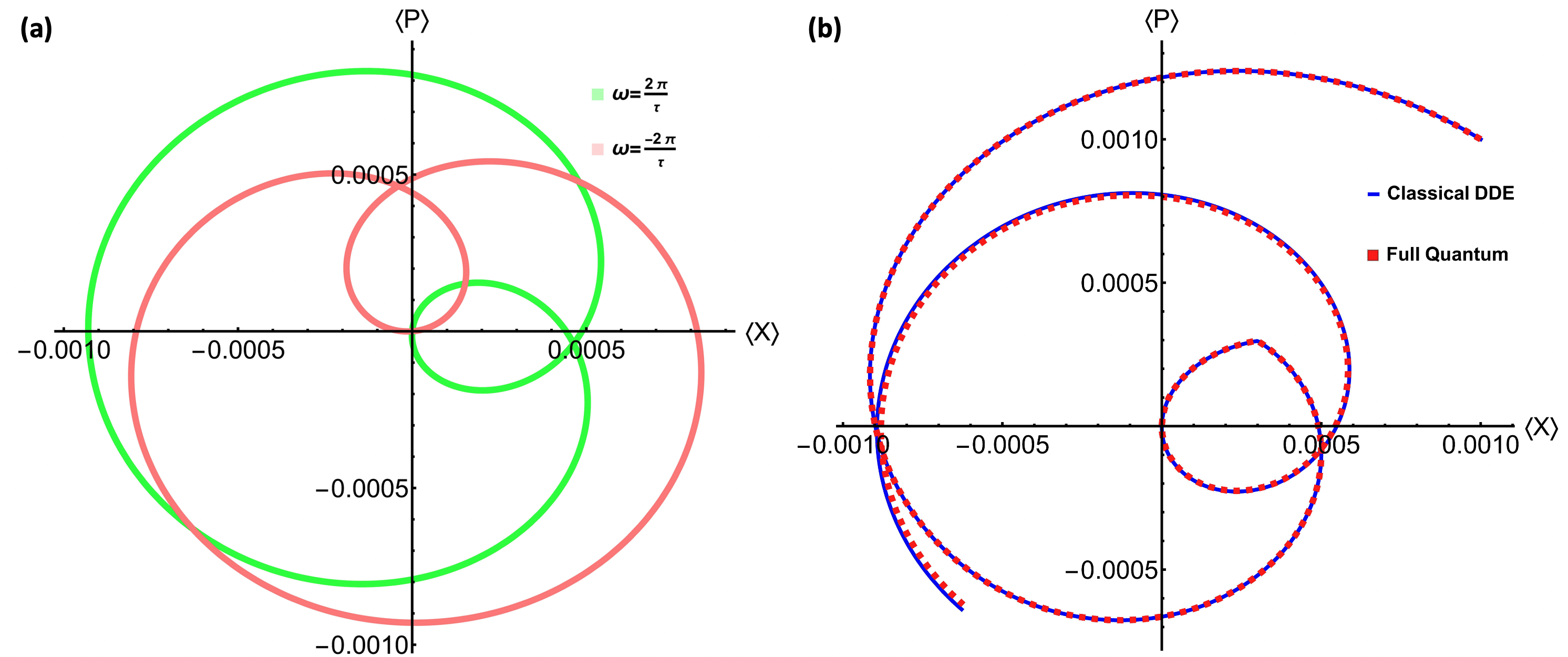}
\caption{Trajectories in phase space for (a) the mean quadratures $(\langle x(t)\rangle, \langle p\rangle)$ by solving their DDEs and (b) comparison of their trajectory when solved using the DDE or by solving the full delayed master equation. We observe very good agreement. For (a) we choose the parameter values $\alpha^\prime \approx -1.2092,\beta^\prime=-2.4184, G=4$, and plot two trajectories with $\omega\tau=2\pi$ and $\omega\tau=-2\pi$, while for (b) we choose $\kappa_1\tau=\kappa_2\tau=1.2092, G=4$
and $\omega\tau=2\pi$ in the quantum system. For the DDE evolution in (b) we specified the past history of the values as $z(t)=z(t=0)*\exp(\gamma t)$, for $t<0$, where $\gamma\sim 10^3\kappa$.}

\label{fig:closedcycles}
\end{figure}

To explore the closed cycles in linear quantum systems, we include an extra Hamiltonian $-\omega\hbar \hat{a}^\dagger \hat{a}$ into the quantum system, which changes the original quantum system described in \Eref{eqn:QuantumDDE} to 
\begin{equation}
\label{eqn:newQuantumDDE}
    \langle \dot{\hat{a}}(t) \rangle=i\omega \langle\hat{a}(t)\rangle-\kappa \langle \hat{a}(t) \rangle-e^{i\phi}\sqrt{G\kappa_1\kappa_2}\,\langle  \hat{a}(t-\tau)\rangle.
\end{equation}
which with the scaled time $\kappa t$ has the form
\begin{equation*}
    \langle \dot{\hat{a}}(t) \rangle=(i\omega/\kappa -1)\langle\hat{a}(t)\rangle-e^{i\phi}\sqrt{G\kappa_1\kappa_2}/\kappa\,\langle  \hat{a}(t-\tau)\rangle.
\end{equation*}
We then simulate our quantum systems using the same set of parameters $\kappa_1=\kappa_2=\kappa, G=4$ and $\omega\tau=2\pi$ as in the classical DDE. We expect to observe the same closed cycles in the quantum case. However, due to the extreme high dimensionality of full quantum cascaded master equation, we are unable to observe the final closed cycle on conventional computers. The excellent agreement between the quantum simulation and the classical DDE depicted in \Fref{fig:closedcycles}-(b) is sufficient evidence to assert that we would observe closed cycles in quantum systems if we could simulate for a greater number of time steps.}

\section{Nonlinear quantum self-oscillators}
We now consider the case when one has a non-linear open-system oscillator which exhibits periodic limit cycles. As mentioned above it is known that for such limit cycles without feedback one typically has phase diffusion due to either thermal or quantum noise. We now add in delayed feedback with gain to see if this limit-cycle associated phase diffusion can be either eliminated or reduced.  From our explorations below we observe that feedback in this case cannot remove phase-diffusion. 

We form such a nonlinear open system by introducing two-photon absorption in the cavity. We first consider  nonlinear damping in the classical DDEs, with the purpose of imposing a threshold oscillation amplitude. One such nonlinear DDE has the following form  \cite{lazarus2015dynamics}:
\begin{equation}
\label{eqn:nonDDE}
    \dot{x}(t)=\alpha x(t)+\beta x(t-\tau)-\gamma_{non} x^3(t),
\end{equation}
where $\gamma_{non}>0$ is the nonlinear damping rate.

\begin{figure}
\centering
\includegraphics[width=\textwidth]{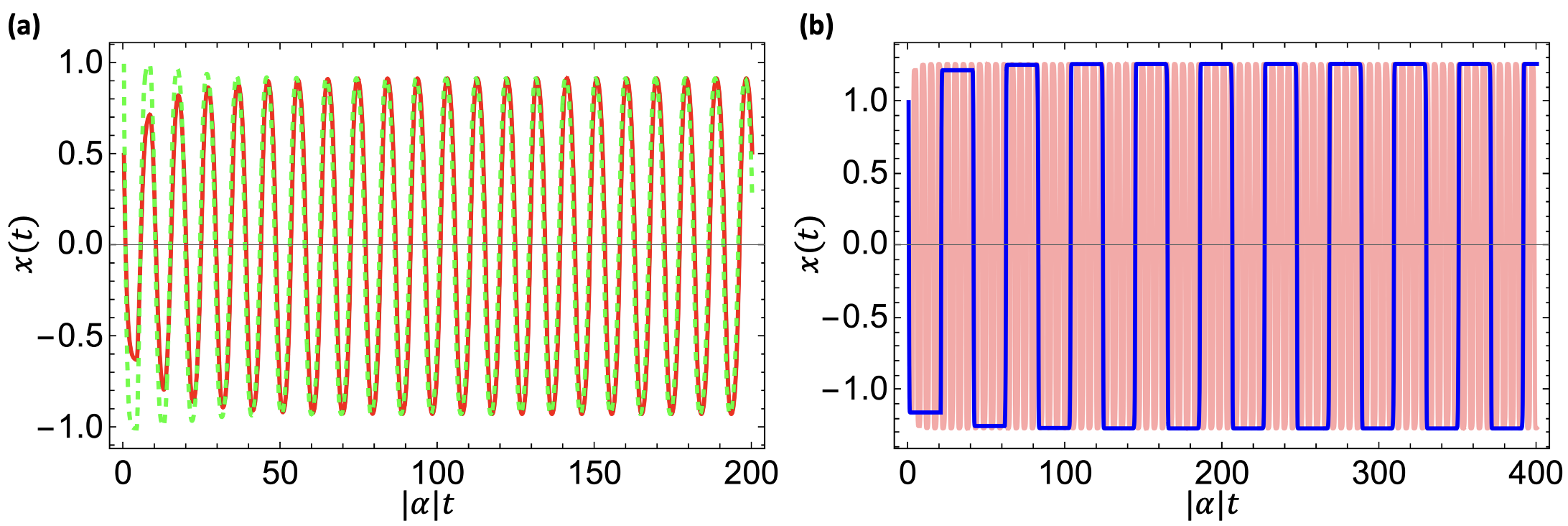}
\caption{Solutions of nonlinear classical DDEs display near-perfect saturated oscillations that are independent of the initial values, and the oscillating wave can be tuned by delay time. (a) Plot of the solution $x(t)$ of nonlinear DDEs starting from two different initial values $x(0)=0.5$ and $x(0)=1$; (b) Plot of the two oscillation waves with different delay times; the red line shows the results with delay time $\kappa\tau=\kappa(\tau_{cr}+1)$ and the blue one shows that of $\kappa\tau=\kappa(\tau_{cr}+10)$.}
\label{fig:5}
\end{figure}

In this nonlinear DDE, we can always find a delay time that is larger than $\tau_{cr}$ such that the solution $x(t)$ of \Eref{eqn:nonDDE} is stably oscillating around an equilibrium point. In \Fref{fig:5}-(a) we plot $x(t)$, the solution to \Eref{eqn:nonDDE}, assuming $x(0)=0.5$ or $1$. \Fref{fig:5}-(b) shows the results of choosing different delay time.

\subsection{Including a two-photon absorption to have a nonlinear quantum DDE}
It would be advantageous if one can obtain a similar nonlinear quantum self-oscillator by introducing a nonlinear damping in the quantum optical system. One of the options to achieve this is to add a two-photon absorption (TPA) \cite{pawlicki2009two} in our ring cavity. TPA is a nonlinear optical phenomenon that occurs when two photons are simultaneously absorbed. This phenomenon has been detected experimentally in many optical materials such as europium-doped crystal \cite{Kaiser1961TWO-PHOTONEu2+}, cesium vapor \cite{Abella1962OpticalVapor}, and Cds \cite{Braunstein1964OpticalCdSdagger}. The TPA used in our setup is to absorb two photons from the cavity $\hat{a}$ at the same time, which will change the master equation of the optical system as:
\begin{equation}
\label{eqn:nonlinear_master_equation}
\eqalign{
        \frac{d\rho}{dt}&=\mathcal{L}_m\rho, t\in [m\tau, (m+1)\tau], \textrm{where}\\
         \mathcal{L}_m\rho&=-\frac{i}{\hbar}\sum_{j=0}^m [H_j,\rho] +\sum_{j=0}^{m}  \bar{N}_{m-j}\kappa_1\left(\mathcal{D}[\hat{a}_j]\rho+\mathcal{D}[\hat{a}_j^\dagger]\rho\right)+\gamma_{non} \sum_{j=0}^m \mathcal{D}[\hat{a}_j^2]\\
          &+(\kappa_1+\kappa_2)\sum_{j=0}^m \mathcal{D}[\hat{a}_j]\rho-\sqrt{G\kappa_1 \kappa_2}\sum_{j=1}^{m} 
         \left\{ [\hat{a}_{j-1}^\dagger,\hat{a}_j\rho]+[\rho \hat{a}_j^\dagger, \hat{a}_{j-1}] \right\}\\           
}
\end{equation}
The DDE of $\langle \hat{a}\rangle $ of this nonlinear system will be:
\begin{equation}
    \label{eqn:nonlinear_QuantumDDE}
    \langle \dot{\hat{a}}(t) \rangle=-\kappa \langle \hat{a}(t) \rangle-e^{i\phi}\sqrt{G\kappa_1\kappa_2}\langle  \hat{a}(t-\tau)\rangle-\gamma_{non} \langle \hat{a}^\dagger \hat{a}\hat{a}\rangle,   
\end{equation}which is similar to the nonlinear DDE \eref{eqn:nonDDE}. However, there is a crucial difference with $\langle \hat{a}^\dagger \hat{a}\hat{a} \rangle \nsim x^3(t)$. This difference will inexorably introduce phase diffusion into the oscillation of $\langle \hat{a} \rangle$ in nonlinear quantum systems.

\begin{figure*}[!ht]
\centering
\includegraphics[width=\textwidth]{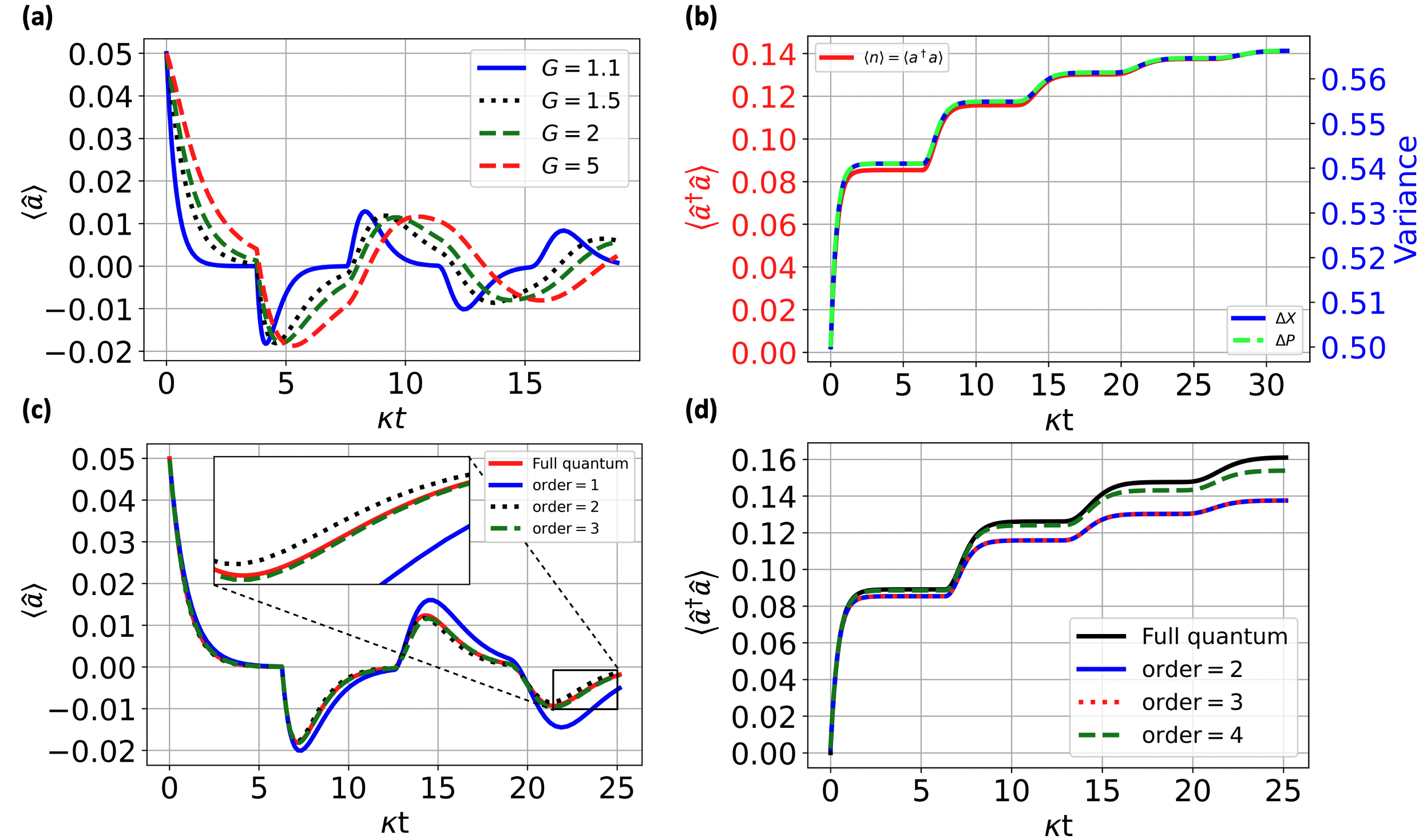}
\caption{Nonlinear self-sustained quantum oscillator shows damped oscillation of $\langle \hat{a} \rangle$, while the energy reaches a steady state bounded value. (a) Behaviour of the expectation value $\langle \hat{a} \rangle$ with different feedback gains $G=(1.1,1.5,2,5)$; (b) Dynamics of the mean photon number $\langle \hat{a}^\dagger \hat{a}\rangle$, $\Delta X(t)$, and $\Delta P(t)$, all of which reach a steady state with time; (c) Plots the convergence of $\langle \hat{a} \rangle$ using the mean-field approach with higher orders of approximation $\textrm{order}=(1,2,3)$, and comparison with the full quantum result, where the inset plot detail is for $\kappa t\in [21,25]$; (d) The convergence of $\langle \hat{a}^\dagger \hat{a} \rangle$ using the mean field approach  with $\textrm{order}=(2,3,4)$ as compared to the full quantum result.}
\label{fig:6}
\end{figure*}

\subsection{Numerical results}
With the master equation \eref{eqn:nonlinear_master_equation}, we can choose $G=1.2, \kappa_1=\kappa_2=\kappa, \gamma_{non}/\kappa=1 $ and any delay time larger than $\tau_{cr}$.
Here we choose $\kappa \tau=\kappa \left(\tau_{cr}+1\right)\approx 6.284 $, to simulate the evolution of $\langle \hat{a} \rangle$. Unlike the results in classical \Eref{eqn:nonDDE}, the evolution of $\langle \hat{a} \rangle$ in the nonlinear quantum system shows dissipative oscillation, primarily due to quantum phase diffusion arising from the $\gamma_{non}\langle \hat{a}^\dagger \hat{a} \hat{a} \rangle$ term.

Due to the computer memory limitations, we can only simulate $4$ time steps in this case. We are still interested in how to simulate over longer times, even though we cannot see a perfect self-oscillation of $\langle \hat{a}\rangle$. Therefore, we also apply a similar mean-field approximation approach to explore the extended time evolution. Like the linear quantum case, we can focus on the ODEs of $\langle \hat{a} \rangle$ instead of the whole master equation, but the nonlinear case will be much more complex due to the nonlinear damping, since ODEs for $\langle \hat{a}^\dagger \hat{a}\hat{a} \rangle $ will be involved with higher order operator products. This results in an infinite number of ODEs involving these higher order operator products. We can truncate these equations by expanding the expectation of higher order operator products in terms of lower order products. This is the original concept of cumulants \cite{Plankensteiner2022QuantumCumulants.jl:Systems,Kubo1962GeneralizedMethod}, and we expect that truncating to higher order quantum correlations will result in more precision in the simulation. The details to apply the cumulants can be found in the \ref{apx:cumulant}.

When we increase to larger time steps, the number of ODEs will increase. However, the resources required is far fewer than those needed to solve the full quantum master equation \eref{eqn:nonlinear_master_equation}. Since the numbers of ODEs required increase rapidly with both time and truncation level, we need to balance the accuracy of this mean-field approximation and the number of ODEs required. We examine the accuracy of this method in \Fref{fig:6}-(c). We observe that convergence of truncation of order $3$ seems excellent. We set the order as $3$ to simulate the effect of feedback gain $G$ on $\langle \hat{a}\rangle$ in \Fref{fig:6}-(a). The results in \Fref{fig:6}-(b), indicate that $\langle \hat{a}^\dagger \hat{a}\rangle$, $\Delta X$ and $\Delta P$ no longer rises in an unbounded fashion but both appear to saturate. We also examine the convergence of photon number of different expansion order to the full quantum master equation in \Fref{fig:6}-(d). Finally, from extensive numerical analysis of this nonlinear SSOs system we could not find conditions which permitted indefinite SSO of $\langle \hat{a}(t)\rangle$ without phase diffusion. 

\section{Conclusions}
In this paper, we have designed the quantum optical self-oscillator, by using time-delay feedback control, both in the linear and nonlinear regimes. We have generalized the standard cascaded theory to the case incorporating amplification, based on which we derived a quantum master equation from the Langevin equation. The linear quantum self-oscillator is composed by an empty ring cavity and a optical amplifier in the feedback loop, which shows perfect oscillation of $\langle\hat{a}(t)\rangle$ without any phase diffusion. For the nonlinear situation, we introduced a two-photon absorption to work as a nonlinear damping, which shows a dissipative oscillation of $\langle\hat{a}(t)\rangle$ as the existed non-delay limit-cycle oscillators. The future work may focus on whether and how we can obtain an ideal quantum nonlinear self-oscillation without phase dissipation by using time-delay feedback.

\appendix
\section{Quantum amplification}
\label{apx:Qamplification}
In this section, we briefly review the theory of quantum amplification. We will require such amplification within the feedback loop. For more details on quantum amplification theory the reader can consult \cite{josse2006universal}.

Optical amplification is inevitably affected by fundamental additional quantum noise no matter whether it is phase sensitive or phase insensitive. From the input-output theory, it is possible to describe a quantum amplification process as:
\begin{equation}
    b_{out}=\sqrt{G}\,b_{in}+\hat{M},
\end{equation}
where $b_{in(out)}$ represents the input (output) annihilation bosonic operators, $G$ is the amplifier gain, and $\hat{M}$ is the operator associated with the additional noise. This additional noise is required to ensure the preservation of the commutation relations $[b_{\bullet},b_{\bullet}^\dagger]=1$, and such noise is intrinsic  even for an ideal quantum amplifier. Following this, the noise operator must satisfy $[\hat{M},\hat{M}^\dagger]=1-G$. If we divide the noise into a fundamental quantum part $\hat{M}_q=\sqrt{G-1}b_{amp}^\dagger$ with $b_{amp}$ being the unavoidable fluctuations of the bosonic mode, and a semi-classical part $\hat{M}_{cl}$. We can obtain the input-output relations of a quantum amplifier working at the quantum noise limit ($\hat{M}_{cl}=0$) as:
\begin{equation}
    b_{out}=\sqrt{G}\,b_{in}+\sqrt{G-1}\,b_{amp}^\dagger.
\end{equation}
The intrinsic quantum noise satisfies the bosonic commutation relation:
\begin{equation}
\eqalign{
    db_{amp}^\dagger db_{amp}=\bar{N}_{amp}dt,\\
    db_{amp}db_{amp}^\dagger=(\bar{N}_{amp}+1)dt,
}
\end{equation}
where $\bar{N}_{amp}$ is the bath occupation of the intrinsic noise.

\section{Calculation details of the master equation}
\label{apx:mastercalculation}
This sections gives the calculation details to obtain the master equation from the coupled Langevin equation, for the cascaded system incorporating amplification.

We show the general cascaded chain and the one incorporates amplifiers in \Fref{fig:cascaded_chain}. Let's use $\hat{o}_i$ to represent any operator for the corresponding $j$-th system in the chain, and their annihilation operators are denoted as $\hat{a}_j$. The dynamics of $\hat{o}_j$ can be written as:
\begin{figure}
  \centering
  \includegraphics[width=\linewidth]{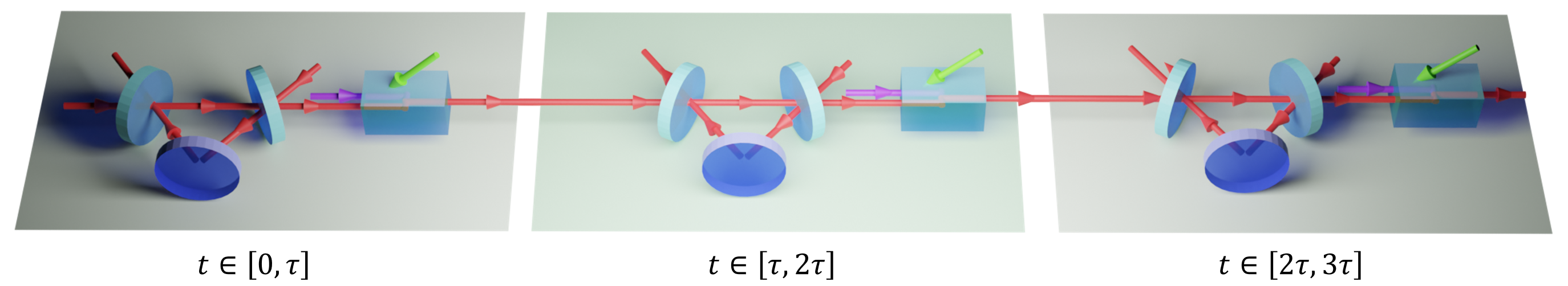}  
\caption{The abstract figure of systems in a cascaded chain; (a) each system is described by a operator $\hat{a}_j$, which is fed in a uni-directional fashion of the previous system; for example, $\hat{a}_2$ is fed by $\hat{a}_1$, $b^{in}_2$ is equal to $b_1^{out}$, and $\hat{a}_3$ is fed by $\hat{a}_2$, etc; (b) the extended cascaded chain with amplifier in each pair of systems, which will introduce noise $b_{amp}$ to the field.}
\label{fig:cascaded_chain}
\end{figure}
For operator $\hat{o}_1$ of the first system:
\begin{equation}
 \label{eqn:cavity1}
\eqalign{
     \dot{\hat{o}}_1=-\frac{i}{\hbar}[\hat{o}_1,H_{sys}(1)]-[\hat{o}_1,\hat{a}_1^\dagger]\left\{\frac{\gamma_1}{2}\hat{a}_1+\sqrt{\gamma_1}b_1^{in}(t)\right\}\cr
     +\left\{\frac{\gamma_1}{2}\hat{a}_1^\dagger+\sqrt{\gamma_1}{b_1^{in}}^\dagger(t)\right\}[\hat{o}_1,\hat{a}_1],\\
    b_1^{out}(t)=\sqrt{\gamma_1}\hat{a}_1(t)+b_1^{in}(t),     
}
\end{equation}

For operator $\hat{o}_2$ of the second system:
\begin{equation}
\eqalign{
 \label{eqn:cavity2}
    \dot{\hat{o}}_2&=-\frac{i}{\hbar}[\hat{o}_2,H_{sys}(1)+H_{sys}(2)]\\
    &-[\hat{o}_2,\hat{a}_1^\dagger]\left\{\frac{\gamma_1}{2}\hat{a}_1+\sqrt{\gamma_1}b_1^{in}(t)\right\}+\left\{\frac{\gamma_1}{2}\hat{a}_1^\dagger+\sqrt{\gamma_1}{b_1^{in}}^\dagger(t)\right\}[\hat{o}_j,\hat{a}_1]\\
    &-[\hat{o}_2,\hat{a}_2^\dagger]\left\{\frac{\gamma_2}{2}\hat{a}_2+\sqrt{\gamma_2}b_{2^\prime}^{in}(t)\right\}+\left\{\frac{\gamma_2}{2}\hat{a}_2^\dagger+\sqrt{\gamma_2}{b_{2^\prime}^{in}}^{\dagger}(t)\right\}[\hat{o}_2,\hat{a}_2]\\
    &-[\hat{o}_2,\hat{a}_2^\dagger]e^{i\phi}\sqrt{G\gamma_1\gamma_2}\hat{a}_1(t)+e^{-i\phi}\sqrt{G\gamma_1\gamma_2}\hat{a}_1^\dagger [\hat{o}_2,\hat{a}_2]\\
     b_2^{in}(t)&=e^{i\phi}\sqrt{G}b_1^{out}(t)+\sqrt{G-1}b_{amp}^\dagger(t)=e^{i\phi}\sqrt{G\gamma_1}\hat{a}_1(t)+b_{2^\prime}^{in}(t),\\
     b_{2^\prime}^{in}(t)&=e^{i\phi}\sqrt{G}b_1^{in}(t)+\sqrt{G-1}b_{amp}^\dagger (t).   
}    
\end{equation}
Similarly, for an operator $\hat{o}_j$ of the $j$-th system:
\begin{equation}
\label{eqn:cavity_any}
\eqalign{
\dot{\hat{o}}_j&=-\frac{i}{\hbar}[\hat{o}_j,H_{sys}(1)+\cdots+H_{sys}(j)]\\
 &-[\hat{o}_j,\hat{a}_1^\dagger]\left\{\frac{\gamma_1}{2}\hat{a}_1+\sqrt{\gamma_1}b_1^{in}(t)\right\}+\left\{\frac{\gamma_1}{2}\hat{a}_1^\dagger+\sqrt{\gamma_1}{b_1^{in}}^\dagger(t)\right\}[\hat{o}_j,\hat{a}_1]\\
 &-[\hat{o}_j,\hat{a}_2^\dagger]\left\{\frac{\gamma_2}{2}\hat{a}_2+\sqrt{\gamma_2}b_{2^\prime}^{in}(t)\right\}+\left\{\frac{\gamma_2}{2}\hat{a}_2^\dagger+\sqrt{\gamma_2}{b_{2^\prime}^{in}}^{\dagger}(t)\right\}[\hat{o}_j,\hat{a}_2]-\cdots \\
 &-[\hat{o}_j,\hat{a}_j^\dagger]\left\{ \frac{\gamma_j}{2}\hat{a}_j+\sqrt{\gamma_j}b_{j^\prime}^{in}(t) \right\}+\left\{\frac{\gamma_j}{2}\hat{a}_j^\dagger+\sqrt{\gamma_j}{b_{j^\prime}^{in}}^{\dagger}(t)\right\}[\hat{o}_j,\hat{a}_j]\\
 &-[\hat{o}_j,\hat{a}_2^\dagger]e^{i\phi}\sqrt{G\gamma_1\gamma_2}\hat{a}_1(t)+e^{-i\phi}\sqrt{G\gamma_1\gamma_2}\hat{a}_1^\dagger [\hat{o}_j,\hat{a}_2]-\cdots\\
 &-[\hat{o}_j,\hat{a}_j^\dagger]e^{i\phi}\sqrt{G\gamma_{j-1}\gamma_j}\hat{a}_{j-1}(t)+e^{-i\phi}\sqrt{G\gamma_{j-1}\gamma_j}\hat{a}_{j-1}^\dagger [\hat{o}_j,\hat{a}_j]. 
}
\end{equation}
Here we use $\gamma_j$ as the decay rate of the $j$-th system, and $b_{amp}$ represents the noise input to the amplifier. All the operators have the form of:
\begin{equation}
    \mathcal{O}_j=\underbrace{I\otimes\cdots \otimes I}_{j\textit{ times}} \otimes \mathcal{O}\otimes \underbrace{I \otimes \cdots \otimes I}_{(m-j-1) \textit{ times}}.
\end{equation}
The operator $\mathcal{O}$ is in the Hilbert space of each individual system, and the operator $\mathcal{O}_j$ resides in the whole Hilbert space which is a tensor product of all the temporally separate systems.

We next derive a valid master equation following the above Heisenberg coupling equations for the cascaded chain. This will be done by first defining various quantum stochastic fields. The most general case which we will consider is that input fields can be written in terms of a quantum white noise part and a coherent part, thus the input noise to the first system can be written as:
\begin{equation}
\label{eqn:inputfield}
\eqalign{
b_1^{in}(t)dt=dB_1(t)+\epsilon_1^{in}(t)dt, \\
dB_1^2(t)=dB_1^{\dagger 2}(t)=0, \\
dB_1(t)dB_1^\dagger(t)=(\bar{N}_1+1)dt, \\
dB_1^\dagger(t)dB_1(t)=\bar{N}_1dt,
}
\end{equation}
while the input field to the second system, which, due to the amplification, will be different. We can write that as follows:
\begin{equation}
\label{eqn:inputfield_j}
\eqalign{
    b_2^{in}(t)dt=dB_1(t)+\epsilon_2^{in}(t)dt,\\
    dB_2^2(t)=dB_2^{\dagger 2}(t)=0,\\
    dB_2(t)dB_2^\dagger(t)=GdB_1(t)dB_1^\dagger(t)+(G-1)db_{amp}^\dagger db_{amp}=(\bar{N}_2+1)dt,\\
    dB_2^\dagger(t)dB_2(t)=GdB_1^\dagger(t)dB_1(t)+(G-1)db_{amp}db_{amp}^\dagger=\bar{N}_2dt.\\
}
\end{equation}
Similarly for the $j$-th system
\begin{equation}
\eqalign{
    b_j^{in}(t)dt=dB_j(t)+\epsilon_j^{in}(t)dt,\\
    dB_j^2(t)=dB_j^{\dagger 2}(t)=0,\\
    dB_j(t)dB_j^\dagger(t)=GdB_{j-1}(t)dB_{j-1}^\dagger(t)+(G-1)db_{amp}^\dagger db_{amp}=(\bar{N}_j+1)dt,\\
    dB_j^\dagger(t)dB_j(t)=GdB_{j-1}^\dagger(t)dB_{j-1}(t)+(G-1)db_{amp}db_{amp}^\dagger=\bar{N}_jdt.
}
\end{equation}

To ensure that all systems in the chain are identical we assume that the input field of the first system $b_1^{in}(t)$ also contains the same amplifier noise, i.e., $\bar{N}_1=G\bar{N}+(G-1)(\bar{N}_{amp}+1)$. Here $\bar{N}$ is the mean photon number of initial input field while $\bar{N}_{amp}$ is that of the amplifier noise. We observe that
\begin{equation}
\label{eqn:meanofinputfield}
\eqalign{
        \bar{N}_1&=G\bar{N}+(G-1)(\bar{N}_{amp}+1),\\
        \bar{N}_2&=G\bar{N}_1+(G-1)(\bar{N}_{amp}+1)=G^2\bar{N}+(G^2-1)(\bar{N}_{amp}+1),\\
        \bar{N}_3&=G\bar{N}_2+(G-1)(\bar{N}_{amp}+1)=G^3\bar{N}+(G^3-1)(\bar{N}_{amp}+1),\\
        & \vdots\\
        \bar{N}_j&=G\bar{N}_{j-1}+(G-1)(\bar{N}_{amp}+1)=G^j\bar{N}+(G^j-1)(\bar{N}_{amp}+1),   }
\end{equation}
meaning that the effective input noise increases with each additional temporal stage from \eref{eqn:meanofinputfield}.

Let's assume $\phi=0$, as this will simplify the choice of $G$ to yield self-sustained oscillations. With these definitions of the stochastic field, the master equation is obtained as 
\begin{equation}
\label{eqnapp:cascaded_master}
\eqalign{
\frac{d\rho_m}{dt}=\frac{i}{\hbar}[\rho,H_{tot}]+\sum_{j=1}^m\left\{\gamma_j (\bar{N}_j+1)\mathcal{D}[\hat{a}_j]\rho+\gamma_j \bar{N}_j\mathcal{D}[\hat{a}_j^\dagger]\rho\right\}\\
+\sum_{j=2}^m\sqrt{G\gamma_{j-1} \gamma_j}\left\{ [\hat{a}_j,\hat{a}_{j-1}\rho]+[\rho \hat{a}_{j-1}^\dagger,\hat{a}_j] \right\}\cr
-\sum_{j=1}^m\left\{\sqrt{\gamma_j}\left[\hat{a}_j^\dagger,\rho  \right]\epsilon_{in}(j,t)-\sqrt{\gamma_j}\epsilon^*_{in}(j,t)\left[\hat{a}_j,\rho\right]\right\}}
\end{equation}
with the initial state at $t=0$ being:
\begin{equation}
    \label{eqnapp:initial_state}
    \rho_m(0)=\underbrace{\rho_S(0)\otimes \cdots \otimes \rho_S(0)}_{m \textit{ times}}.
\end{equation}

\section{Cumulant theory to expand the ODES}
\label{apx:cumulant}
We make use of the joint cumulant formula to expand the operator products of order $n$ as \cite{Plankensteiner2022QuantumCumulants.jl:Systems}:
\begin{equation}
\label{eqn:joint_cumulants}
    \langle X_1X_2\cdots X_n \rangle=\sum_{p\in \mathcal{P}(\mathcal{I})\setminus \mathcal{I}} (|p|-1)\!(-1)^{|p|}\prod_{B\in p}\left\langle \prod_{i\in B}X_i \right\rangle
\end{equation}
Here $\mathcal{I} = \{1,2,...,n\}$, $P(\mathcal{I})$ is the set of all partitions of $\mathcal{I}$, $|p|$ denotes the length of the partition $p$, and $B$ runs over the blocks of each partition. We will give two examples here to show the partitions, when $n=2$ we only keep the operator products of order $1$ as:
\begin{equation}
\langle X_1X_2 \rangle=\langle X_1 \rangle\langle X_2 \rangle,\nonumber\\
\end{equation}
Similarly when $n=3$, we expand all operator products of order $3$ and higher as:
\begin{equation}
\label{eqn:examples_expansion}
\eqalign{\langle X_1X_2X_3 \rangle &=\langle X_1 X_2\rangle\langle X_3\rangle+\langle X_1X_2 \rangle\langle X_3\rangle+\langle X_1 \rangle\langle X_2X_3\rangle\cr
&-2\langle X_1\rangle\langle X_2\rangle\langle X_3\rangle}.
\end{equation}

Therefore, when we calculate the ODE of $\langle \hat{a}_0 \rangle$ in each time interval based on the nonlinear master equation, the above joint cumulant \Eref{eqn:joint_cumulants} will be used to expand the expectation of nonlinear operator products of a given order. This permits us to carefully truncate the set of ODEs to a given level of cumulant and this is often know as a ``cumulant expansion". As an example, we show the expansion to second order over the first $2$ time steps.
\begin{equation}
    \label{eqn:nonlinear_mean_first}
\eqalign{
    t\in [0,\tau], \\
         \langle \dot{\hat{a}}_0 \rangle =-\kappa \langle\hat{a}_0 \rangle-\gamma \langle \hat{a}_0^\dagger \hat{a}_0\hat{a}_0 \rangle \\
        =-\kappa \langle\hat{a}_0 \rangle-\gamma\left(2\langle \hat{a}_0^\dagger \hat{a}_0 \rangle\langle \hat{a}_0\rangle+\langle \hat{a}_0^\dagger \rangle\langle \hat{a}_0\hat{a}_0 \rangle-2\langle \hat{a}_0^\dagger \rangle\langle\hat{a}_0\rangle\langle\hat{a}_0\rangle\right)\\
        \dot{\langle\hat{a}_0^\dagger \hat{a}_0\rangle} \,=-2\kappa \langle \hat{a}_0^\dagger \hat{a}_0 \rangle-2\gamma \langle \hat{a}_0^\dagger \hat{a}_0^\dagger \hat{a}_0\hat{a}_0 \rangle \\
        =-2\kappa \langle \hat{a}_0^\dagger \hat{a}_0 \rangle-2\gamma \left(\langle \hat{a}_0^\dagger \hat{a}_0^\dagger \rangle\langle \hat{a}_0\hat{a}_0 \rangle+2\langle \hat{a}_0^\dagger \hat{a}_0 \rangle^2-2\langle\hat{a}_0\rangle\langle\hat{a}_0\rangle\langle\hat{a}_0^\dagger\rangle\langle\hat{a}_0^\dagger\rangle \right)\\
         \dot{\langle \hat{a}_0 \hat{a}_0   \rangle}\, =-(2\kappa+\gamma) \langle \hat{a}_0 \hat{a}_0  \rangle-2\gamma \langle \hat{a}_0^\dagger \hat{a}_0\hat{a}_0\hat{a}_0 \rangle \\
        =-(2\kappa+\gamma) \langle \hat{a}_0 \hat{a}_0  \rangle-2\gamma\left(3\langle\hat{a}_0^\dagger \hat{a}_0\rangle \langle\hat{a}_0 \rangle\langle\hat{a}_0 \rangle-2\langle\hat{a}_0^\dagger\rangle\langle\hat{a}_0\rangle\langle\hat{a}_0\rangle\langle\hat{a}_0\rangle\right)      
}
\end{equation}

\begin{equation}
    \label{eqn:nonlinear_mean_second}
\eqalign{
t\in[\tau,2\tau],\\
        \langle \dot{\hat{a}}_0 \rangle=-\kappa \langle\hat{a}_0 \rangle-\gamma\left(2\langle \hat{a}_0^\dagger \hat{a}_0 \rangle\langle \hat{a}_0\rangle+\langle \hat{a}_0^\dagger \rangle\langle \hat{a}_0\hat{a}_0 \rangle-2\langle \hat{a}_0^\dagger \rangle\langle\hat{a}_0\rangle\langle\hat{a}_0\rangle\right)-\eta \langle\hat{a}_1\rangle\\
        \langle\dot{\hat{a}}_1\rangle=-\kappa \langle\hat{a}_1\rangle-\gamma \langle \hat{a}_1^\dagger \hat{a}_1\hat{a}_1 \rangle \\
        =-\kappa \langle\hat{a}_1\rangle-\gamma \left( 2\langle \hat{a}_1^\dagger \hat{a}_1 \rangle\langle \hat{a}_1\rangle+\langle \hat{a}_1^\dagger \rangle\langle \hat{a}_1\hat{a}_1 \rangle-2\langle \hat{a}_1^\dagger \rangle\langle\hat{a}_1\rangle\langle\hat{a}_1\rangle \right) \\
        \dot{\langle \hat{a}_0^\dagger \hat{a}_0\rangle}\, =-2\kappa \langle \hat{a}_0^\dagger \hat{a}_0 \rangle-2\gamma \left(\langle \hat{a}_0^\dagger \hat{a}_0^\dagger \rangle\langle \hat{a}_0\hat{a}_0 \rangle+2\langle \hat{a}_0^\dagger \hat{a}_0 \rangle^2-2\langle\hat{a}_0\rangle\langle\hat{a}_0\rangle\langle\hat{a}_0^\dagger\rangle\langle\hat{a}_0^\dagger\rangle \right) \\
        -\eta \langle \hat{a}_0^\dagger \hat{a}_1 \rangle-\eta \langle \hat{a}_1^\dagger \hat{a}_0 \rangle \\
        \dot{\langle \hat{a}_0 \hat{a}_0 \rangle}\, =-(2\kappa+\gamma) \langle \hat{a}_0 \hat{a}_0  \rangle-2\gamma\left( 3\langle\hat{a}_0^\dagger \hat{a}_0\rangle\langle\hat{a}_0\hat{a}_0\rangle-2\langle\hat{a}_0\rangle\langle\hat{a}_0\rangle\langle\hat{a}_0\rangle\langle\hat{a}_0^\dagger\rangle \right) \\
        \dot{\langle \hat{a}_0\hat{a}_1 \rangle}\,=-2\kappa \langle \hat{a}_0\hat{a}_1 \rangle -\eta\langle \hat{a}_1^\dagger \hat{a}_1 \rangle-\gamma ( \langle \hat{a}_0^\dagger\hat{a}_1 \rangle\langle\hat{a}_0\hat{a}_0\rangle+2\langle\hat{a}_0\hat{a}_1\rangle \langle \hat{a}_0^\dagger \hat{a}_0\rangle\\
        +\langle \hat{a}_1^\dagger \hat{a}_0 \rangle\langle \hat{a}_1\hat{a}_1 \rangle \nonumber-2\langle\hat{a}_1\rangle \langle\hat{a}_0^\dagger\rangle \langle\hat{a}_0\rangle \langle \hat{a}_0\rangle+2\langle\hat{a}_1^\dagger \hat{a}_1\rangle \langle\hat{a}_1\hat{a}_0 \rangle\\
        -2\langle\hat{a}_1^\dagger \rangle \langle \hat{a}_1 \rangle \langle\hat{a}_1\rangle \langle\hat{a}_0\rangle)\\
        \dot{\langle \hat{a}_0^\dagger \hat{a}_1 \rangle}\,=-2\kappa \langle \hat{a}_0^\dagger \hat{a}_1 \rangle-\eta \langle \hat{a}_1^\dagger \hat{a}_1 \rangle-\gamma (2\langle \hat{a}_0^\dagger\hat{a}_1\rangle \langle \hat{a}_0^\dagger \hat{a}_0 \rangle+\langle\hat{a}_0\hat{a}_1\rangle\langle \hat{a}_0^\dagger \hat{a}_0^\dagger \rangle \\
        -2\langle\hat{a}_1 \rangle\langle\hat{a}_0^\dagger\rangle\langle\hat{a}_0^\dagger\rangle\langle\hat{a}_0\rangle +\langle \hat{a}_1^\dagger \hat{a}_0 \rangle\langle \hat{a}_1\hat{a}_1 \rangle \\
        +2\langle \hat{a}_1^\dagger \hat{a}_1 \rangle \langle\hat{a}_1\hat{a}_0^\dagger \rangle+\langle\hat{a}_1^\dagger \hat{a}_0^\dagger\rangle\langle\hat{a}_1\hat{a}_1\rangle -2\langle\hat{a}_1^\dagger\rangle\langle \hat{a}_1\rangle\langle\hat{a}_1\rangle \langle\hat{a}_0^\dagger \rangle)\\
        \dot{\langle \hat{a}_1\hat{a}_1 \rangle}\,=-(2\kappa +\gamma)\langle \hat{a}_1\hat{a}_1 \rangle-2\gamma \left( 3\langle\hat{a}_1^\dagger \hat{a}_1\rangle \langle \hat{a}_1\hat{a}_1 \rangle-2\langle\hat{a}_1^\dagger\rangle\langle\hat{a}_1\rangle\langle\hat{a}_1\rangle\langle\hat{a}_1\rangle \right)\\
        \dot{\langle \hat{a}_1^\dagger \hat{a}_1 \rangle}\,=-2\kappa \langle \hat{a}_1^\dagger \hat{a}_1 \rangle-2\gamma \left( \langle\hat{a}_1^\dagger \hat{a}_1^\dagger\rangle \langle \hat{a}_1\hat{a}_1 \rangle+2\langle \hat{a}_1^\dagger \hat{a}_1 \rangle^2-2\langle\hat{a}_1^\dagger\rangle\langle\hat{a}_1^\dagger\rangle\langle\hat{a}_1\rangle \langle\hat{a}_1\rangle \right)\\
}
\end{equation}

\bibliographystyle{iopart-num}
\bibliography{Quantum_self-oscillation_with_time_delay_feedback_arXiv}
\label{sec:refs}

\end{document}